\documentclass[aps,prd,twocolumn,superscriptaddress,amssymb,amsmath,nofootinbib]{revtex4}

\usepackage{color}
\usepackage{graphicx}
\usepackage{float}
\usepackage[usenames,dvipsnames]{xcolor}
\usepackage{multirow}
\usepackage[normalem]{ulem}
\usepackage{slashed}

\begin{document}

\preprint{}

\title{Di-Higgs and tri-Higgs boson signals of muon $g-2$ at a muon collider}

\author{Radovan Dermisek}

\email[]{dermisek@indiana.edu}

\author{Keith Hermanek}

\email[]{khermane@iu.edu}

\affiliation{Physics Department, Indiana University, Bloomington, IN, 47405, USA}

\author{Navin McGinnis}

\email[]{nmcginnis@triumf.ca}

\affiliation{TRIUMF, 4004 Westbrook Mall, Vancouver, BC, Canada V6T 2A3}


\date{October 11, 2021}

\begin{abstract}
We show that new physics explanations of the muon $g-2$ anomaly by the contributions of new leptons mediated by the standard model Higgs boson necessarily lead to large rates for $\mu^+ \mu^- \to hh$ and $\mu^+ \mu^- \to hhh$ irrespectively of details of the model or the scale of new physics. For new leptons with the same quantum numbers as the standard model leptons, cross sections are expected to be about 240 ab for $\mu^+ \mu^- \to hh$ independently of the center of mass energy, $\sqrt{s}$, and about 2.7 ab for $\mu^+ \mu^- \to hhh$ for $\sqrt{s} = 1$ TeV and growing quadratically with $\sqrt{s}$. Predictions for models featuring new leptons with different quantum numbers and for a type-II two Higgs doublet model, where additional Higgs bosons can contribute to  muon $g-2$, are also presented.

\end{abstract}


\maketitle

\section{Introduction}

The measured value of the muon anomalous magnetic moment, $a_\mu = (g-2)_{\mu}/2$, represents one of the most significant deviations from predictions of the standard model (SM). Among the simplest new physics explanations are scenarios with additional contributions,  $\Delta a_{\mu}$, from new leptons mediated by SM gauge and Higgs bosons~\cite{Kannike:2011ng, Dermisek:2013gta, Poh:2017tfo,Dermisek:2021ajd, Arkani-Hamed:2021xlp}. The possible mass enhancement in these contributions allows for very heavy leptons,  far beyond the reach of future colliders, and thus the confirmation of such explanations might solely rely on indirect evidence. 

Such indirect evidence includes modifications of muon couplings to $Z$, $W$ and Higgs bosons from SM predictions. Although the required modifications of $Z$ and $W$ couplings can be fully explored at future colliders~\cite{Dermisek:2021ajd}, they are  not tightly related to $(g-2)_{\mu}$. On the other hand, the modification of muon Yukawa coupling is directly related to the explanation of $(g-2)_{\mu}$ and could be observed at the Large Hadron Collider (LHC)~\cite{Kannike:2011ng, Dermisek:2013gta, Freitas:2014pua, Poh:2017tfo, Dermisek:2021ajd}. However, it turns out that for new leptons with the same quantum numbers as SM leptons, the one sigma range of $(g-2)_{\mu}$ is consistent with the SM prediction for $h\to \mu^+ \mu^-$, and exactly the SM value is predicted for $(g-2)_{\mu}$ just slightly below the current central value~\cite{Dermisek:2021ajd}. Thus, the LHC or any future collider might only find an evidence for this explanation, if $h\to \mu^+ \mu^-$ deviates from the SM prediction, but will not be able to rule it out. 
 
\begin{figure}[t]
\includegraphics[scale=0.29]{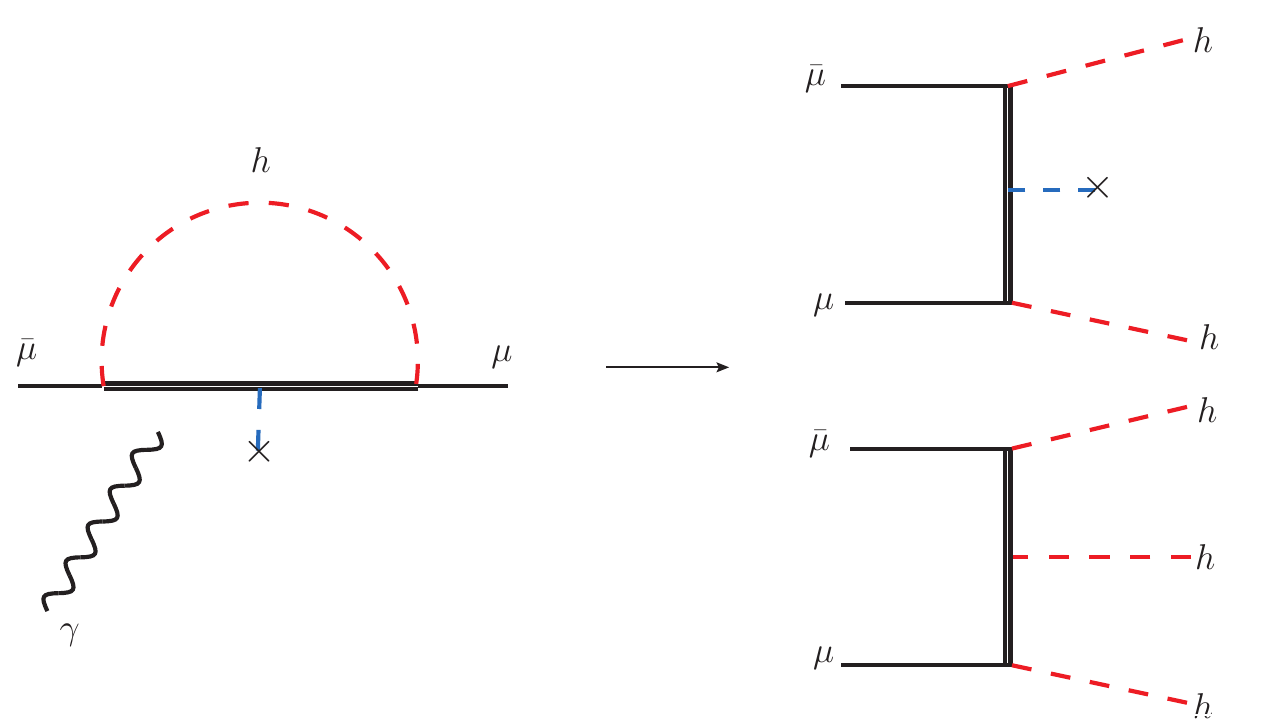}
\caption{Mass enhanced Higgs boson contribution to $(g-2)_{\mu}$ is related to $\mu^+\mu^- \to hh$ by removing the photon and cutting the loop and  $\mu^+\mu^- \to hhh$ by also replacing the vev with $h$.}
\label{fig:g-2_multi-higgs}
\end{figure}

In this paper we show that there are unavoidable signals of contributions to $\Delta a_{\mu}$ from new leptons mediated by the SM Higgs boson at muon colliders. Production cross sections $\mu^+\mu^- \to hh$ and $\mu^+\mu^- \to hhh$ are predicted by $\Delta a_\mu$ without a free parameter in the limit of heavy new leptons, see Fig.~\ref{fig:g-2_multi-higgs}.  For new leptons with the same quantum numbers as SM leptons, cross sections are expected to be about 240 ab for $\mu^+ \mu^- \to hh$ independently of the center of mass energy, $\sqrt{s}$, and about 2.7 ab for $\mu^+ \mu^- \to hhh$ for $\sqrt{s} = 1$ TeV and growing quadratically with $\sqrt{s}$. This presents a unique discovery opportunity, where the scenario could be easily tested even at  low energies.    
Since these predictions depend only on the quantum numbers of new leptons, measuring these rates can point to the correct model without observing the new leptons directly.

Furthermore, we show that in models where SM Higgs and gauge bosons provide only a partial contribution to $\Delta a_{\mu}$, measuring these rates provides information on the parameter that controls the ratio of the SM Higgs and other contributions. For example, in a two Higgs doublet model (2HDM), 
it is the ratio of vacuum expectation values (vevs) of the two Higgs doublets, $\tan \beta$,  that controls the contribution of heavy Higgses. We discuss related mixed di-Higgs and tri-Higgs signals in 2HDM and also signals in models  with new mediators not participating in electroweak symmetry breaking (EWSB).

\section{Brief Summary of a Model}

We start with an extension of the lepton sector of the SM model by vectorlike pairs of SU(2) doublets, $L_{L,R}$, and SU(2) singlets, $E_{L,R}$, where $L_L$ and $E_R$ have the same quantum numbers as SM leptons. For the discussion of $(g-2)_{\mu}$ only mixing of the 2nd generation with new leptons is relevant. This can be enforced by individual lepton number conservation, or, if mixing with all three generations is present, it is assumed that the relevant couplings are sufficiently small to satisfy constraints from flavor changing processes and can be neglected. The most general Lagrangian consistent with these assumptions contains, in addition to kinetic terms, the following Yukawa and mass terms: 
\begin{flalign}
\mathcal{L}\supset& - y_{\mu}\bar{l}_L\mu_{R}H - \lambda_{E}\bar{l}_{L}E_{R}H  - \lambda_{L}\bar{L}_{L}\mu_{R}H  - \lambda\bar{L}_{L}E_{R}H  \nonumber \\   
&   - \bar{\lambda}H^{\dagger}\bar{E}_{L}L_{R}  - M_{L}\bar{L}_{L}L_{R} - M_{E}\bar{E}_{L}E_{R}  + h.c.,
\label{eq:lagrangian}	
\end{flalign}
where doublet components of leptons are labeled as: $l_{L}=( \nu_{\mu},  \mu_{L} )^T$ and  $ L_{L,R}= ( L_{L,R}^{0}, L_{L,R}^{-})^T$. 
In the process of EWSB the Higgs field develops a vev,
\begin{equation}
\hspace{0.25cm} H = \left( \begin{matrix} 0 \\ v + \frac{1}{\sqrt{2}} h \end{matrix} \right),  \hspace{0.3cm} v = 174 \; {\rm GeV},
\end{equation}
where $h$ is the SM Higgs boson, and the charged lepton mass matrix  is generated:
\begin{equation}
	(\bar{\mu}_{L}, \bar{L}_{L}^{-}, \bar{E}_{L})\begin{pmatrix}y_{\mu}v&0 &\lambda_{E}v\\\lambda_{L}v&M_{L}&\lambda v\\0&\bar{\lambda}v& M_{E}\end{pmatrix}\begin{pmatrix}\mu_{R}\\ L_{R}^{-}\\ E_{R}\end{pmatrix}.
\label{eq:mass_matrix}
\end{equation}
Diagonalizing this matrix leads to two new mass eigenstates, $e_4$ and $e_5$, with masses 
$\simeq M_{L,E}$. 
As a result of mixing, couplings between the muon and heavy leptons are generated, and couplings of the muon to $W$, $Z$, and $h$ are modified from their SM values. General formulas  and useful approximate formulas can be found in \cite{Dermisek:2013gta,Dermisek:2015oja,Dermisek:2021ajd}.

\section{Di-Higgs and Tri-Higgs signals}

\begin{figure}[b]
\includegraphics[scale=0.29]{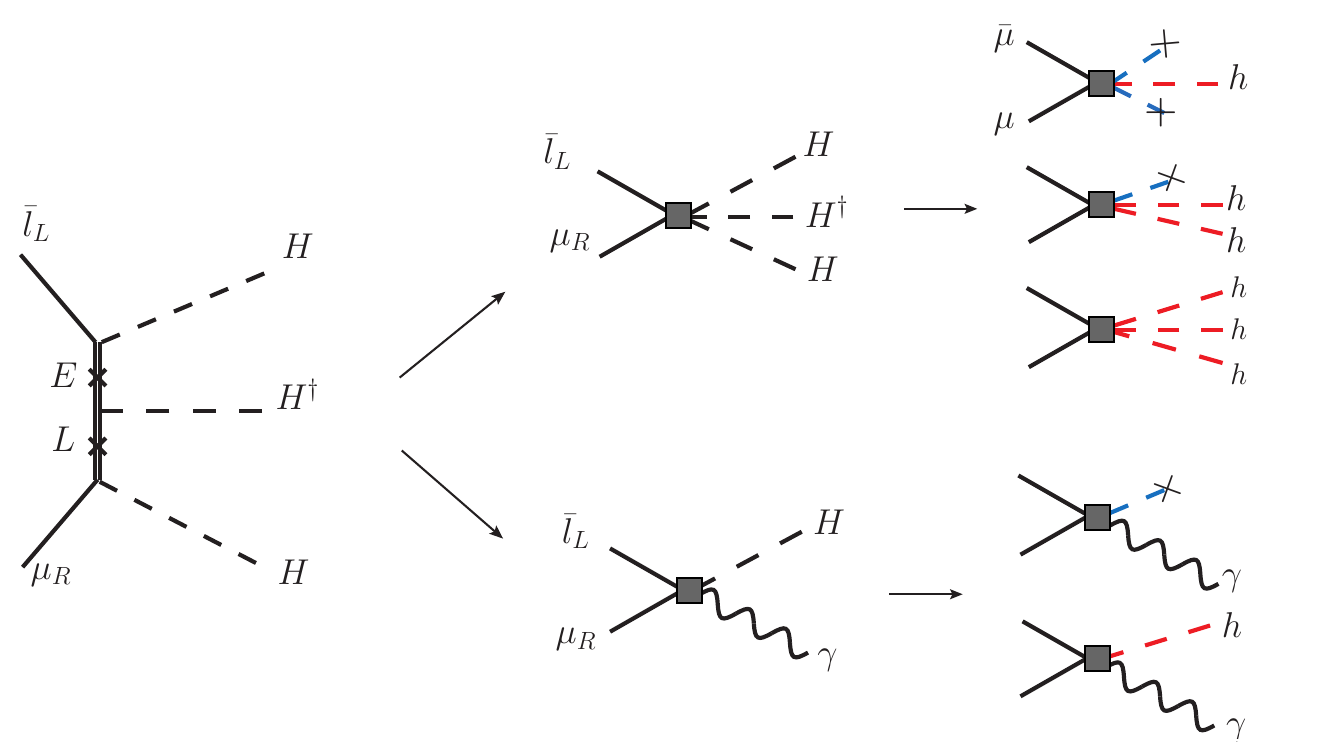}
\caption{Tree level diagram  generating the effective $\bar{l}_L\mu_{R}H H^\dagger H$ operator related to  $(g-2)_{\mu}$. It contributes to the muon mass, muon Yukawa coupling, and creates $\bar{\mu}\mu h^2$ and $\bar{\mu}\mu h^3$ couplings.}
\label{fig:diags}
\end{figure}

At energy scales much below $M_{L,E} $  the Lagrangian in Eq.~(\ref{eq:lagrangian}) reduces to the muon Yukawa term and  dimension-6 operator:
\begin{equation}
 \mathcal{L}\supset - y_{\mu}\bar{l}_L\mu_{R}H - \frac{\lambda_L \bar \lambda \lambda_E}{M_L M_E}\bar{l}_L\mu_{R}H H^\dagger H + h.c..
 \label{eq:eff_lagrangian}
\end{equation}
This operator is a new source of the muon mass and thus possibly modifies $h\to \mu^+\mu^-$. It is  also directly linked to the modification of $(g-2)_{\mu}$, and it  leads to $\mu^+\mu^- \to hh$ and $\mu^+\mu^- \to hhh$, see Fig.~\ref{fig:diags}. The dimension-6 operators arising from kinetic terms of heavy fields $C_{R}(\bar{\mu}_{R}H^{\dagger})i\slashed{D}(\mu_{R}H)$ and $C_{L}(\bar{l}_{L}H)i\slashed{D}(l_{L}H^{\dagger})$, where $C_{L}=\lambda_{E}^{2}/M_{E}^{2}$ and $C_{R}=\lambda_{L}^{2}/M_{L}^{2}$, will also contribute to $\mu^+\mu^- \to hh$. The operators with derivatives acting on muon fields can be rewritten, using equations of motion, in the form of the dimension-6 operator in Eq.~\ref{eq:eff_lagrangian}. The corresponding contributions, however, will be proportional to $y_{\mu}$ and thus suppressed by $m_{\mu}/v$ assuming order one couplings. Similarly, the operators with derivatives acting on Higgs fields can be rewritten as operators with derivatives acting on the muon fields (via integration by parts up to surface terms), and thus will also be suppressed by $m_{\mu}/v$.

Inserting a vev for every Higgs field in the effective Lagrangian the muon mass becomes
\begin{equation}
m_\mu = y_\mu v + m_\mu^{LE} ,
\label{eq:m_mu}
\end{equation}
where
\begin{equation}
m_\mu^{LE} \equiv \frac{\lambda_{L} \bar{\lambda} \lambda_{E}}{M_{L}M_{E}} v^3  
\label{eq:m^LE}
\end{equation}
is the contribution to  muon mass from dimension-6 operator. 

The contribution to muon anomalous magnetic moment, assuming $M_{L,E} \gg M_{EW}$, is given by~\cite{Kannike:2011ng, Dermisek:2013gta}
\begin{equation}
\Delta a_{\mu} 
= - \frac{1}{16\pi^{2}}  \frac{m_\mu m_\mu^{LE}}{v^2}.
\label{eq:dela}
\end{equation}
It is directly related to $m_\mu^{LE}$ and the explanation of the  4.2$\sigma$  deviation of the measured value from the SM prediction~\cite{Abi:2021gix, Aoyama:2020ynm},
\begin{equation}
\Delta a^{exp}_{\mu}\equiv a^{exp}_{\mu} - a_{\mu}^{SM} = (2.51 \pm 0.59)\times 10^{-9},
\end{equation}
requires
\begin{equation}
m_\mu^{LE}/m_\mu = -1.07 \pm 0.25.
\end{equation}
Thus, the effective Lagrangian (\ref{eq:eff_lagrangian}) is completely fixed by the muon mass and muon $g-2$, and predictions for all other observables resulting from it are unique.

Interactions of the muon with the SM Higgs boson are described by the following Lagrangian:
%
%
\begin{equation}
\mathcal{L}\supset - \frac{1}{\sqrt{2}} \, \lambda^h_{\mu\mu}\, \bar{\mu}\mu h - \frac{1}{2} \, \lambda^{hh}_{\mu\mu}\, \bar{\mu}\mu h^2 - \frac{1}{3!}\, \lambda^{hhh}_{\mu\mu}\, \bar{\mu}\mu h^3,
\label{eq:lagrangian_h}	
\end{equation}
where $\mu$ is the Dirac spinor containing $\mu_{L,R}$.
The muon Yukawa coupling is obtained from terms in Eq.~(\ref{eq:eff_lagrangian}) with one $h$:
\begin{equation}
\lambda^h_{\mu \mu} = y_\mu + 3 m_\mu^{LE}/v = ( m_\mu +2 m_\mu^{LE})/v.
\label{eq:lambda_h}
\end{equation}
Clearly the SM prediction for $h \to \mu\mu$ rate can be dramatically altered by $m_\mu^{LE}$. 
Note however, that both $m_\mu^{LE} = 0$ and $m_\mu^{LE} = -m_\mu$ lead to exactly the SM prediction for $h \to \mu\mu$, and, amusingly, the central value of $\Delta a_\mu$ requires $m_\mu^{LE} \simeq -m_\mu$. The one sigma range of $\Delta a_\mu$ suggests that
\begin{equation}
R_{h\to \mu^+\mu^-} \equiv  \frac{BR(h \to \mu^+\mu^-)}{BR(h \to \mu^+\mu^-)_{SM}} = \left( 1+ 2\frac{m_\mu^{LE}}{m_\mu}\right)^2
\label{eq:R}
\end{equation}
is within the range
\begin{equation}
R_{h\to \mu^+\mu^-} = 1.32^{+1.40}_{-0.90}.
\end{equation}
The current upper limit is 2.2~\cite{Aad:2020xfq}. 

The $\bar{\mu}\mu h^2$ and $\bar{\mu}\mu h^3$ couplings directly follow from the dimension-6 operator:
\begin{eqnarray}
\lambda^{hh}_{\mu \mu} &=&  3 \; m_\mu^{LE}/v^2 \label{eq:lambda_hh},\\
\lambda^{hhh}_{\mu \mu} &=& \frac{3}{\sqrt{2}} \;m_\mu^{LE}/v^3 \label{eq:lambda_hhh}, 
\end{eqnarray}
and they are also fixed by $\Delta a_\mu$. Obvious possible manifestations of these couplings are di-Higgs and tri-Higgs, $\mu^+\mu^- \to hh$ and $hhh$, productions at a muon collider. Corrections to these couplings from other dimension-6 operators mentioned above will be proportional to $y_{\mu}vC_{L,R}$ for di-Higgs and $y_{\mu}v^{2}C_{L,R}$ for tri-Higgs.

\begin{figure}[t]
\includegraphics[scale=0.25]{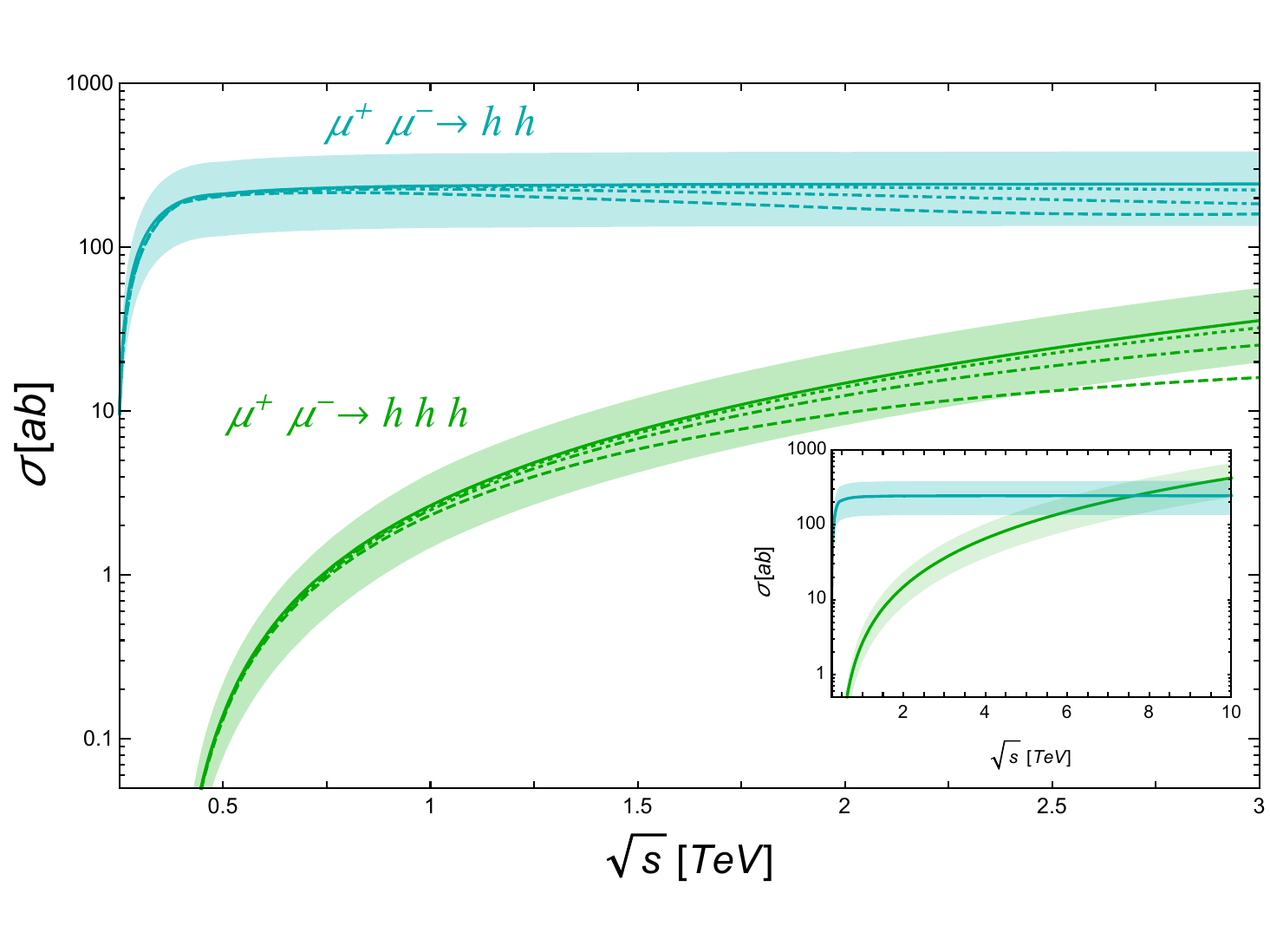}
\caption{Total cross sections for $\mu^+ \mu^- \to hh$ and $\mu^+ \mu^- \to hhh$ as functions of $\sqrt{s}$ corresponding to the  central value of $\Delta a_\mu$ (solid lines) and its  one sigma range (shaded regions) from the effective lagrangian. Dashed, dashed-dotted, and dotted lines indicate predictions of the full model for $M_{L,E} = 3,\,5,\,10$ TeV with all $\lambda$-couplings set equal and fixed to explain the central value of $\Delta a_\mu$. The inset extends the range of $\sqrt{s}$ to 10 TeV.}
\label{fig:cross_sections_SM}
\end{figure}

The total cross sections for these processes, neglecting the muon mass and the Higgs mass,  are as follows:
\begin{eqnarray}
\sigma_{\mu^+\mu^- \to hh} &=&  \frac{\left|\lambda^{hh}_{\mu \mu}\right|^2}{64 \pi}   = \frac{9}{64 \pi} \left(\frac{m_\mu^{LE}}{v^2}\right)^2 \label{eq:sigma_hh},\label{eq:EFT_xsections_1}\\
\sigma_{\mu^+\mu^- \to hhh}  &=& \frac{\left|\lambda^{hhh}_{\mu \mu}\right|^2}{6144 \pi^3} s = \frac{3}{4096 \pi^3} \left(\frac{m_\mu^{LE}}{v^3}\right)^2 s \label{eq:sigma_hhh}.
\label{eq:EFT_xsections_2}
\end{eqnarray}
These formulas are excellent approximations to cross sections well above the production thresholds. Both cross sections, calculated from the effective lagrangian implemented in {\tt FeynRules}~\cite{Degrande:2011ua} using {\tt MadGraph5}~\cite{Alwall:2014hca}, are plotted as functions of $\sqrt{s}$ in Fig.~\ref{fig:cross_sections_SM} for $m_\mu^{LE}$ fixed by 
$\Delta a_\mu$. 
We see that the total cross section for  $\mu^+ \mu^- \to hh$, away from the threshold, is about 240 ab independently of the center of mass energy. For $\mu^+ \mu^- \to hhh$ it is about 2.7 ab 
for $\sqrt{s} = 1$ TeV and growing quadratically with $\sqrt{s}$. The two cross sections are equal for $\sqrt{s} \simeq 7.5$ TeV.

The equalities in Eqs.~(\ref{eq:m_mu}), (\ref{eq:dela}), (\ref{eq:lambda_h}), (\ref{eq:R}) and (\ref{eq:lambda_hh})-(\ref{eq:sigma_hhh}) should be replaced by $\simeq$ if the new leptons are not very heavy and the effective theory is not a good approximation. For completeness, in Fig.~\ref{fig:cross_sections_SM}  we also plot predictions of the full model for $M_{L,E} = 3,\,5,\,10$ TeV with all $\lambda$-couplings set equal and fixed to explain the central value of $\Delta a_\mu$. 
The formulas for  $\sigma_{\mu^+\mu^- \to hh}$  and $\sigma_{\mu^+\mu^- \to hhh}$ in the full model are presented in the appendix. 
 We see that even at $\sqrt{s} = M_{L,E} $ the results based on effective lagrangian are good approximations to predictions of the full model.

\section{Predictions in 2HDM}

The above formulas  can be straightforwardly modified for the type-II 2HDM version of the model, where the structure of Yukawa couplings of SM and new leptons  to two Higgs doublets, $H_u$ and $H_d$, follows from the $Z_2$ symmetry. Since charged leptons couple to $H_d$, the $H$ in both Lagrangians (\ref{eq:lagrangian}) and (\ref{eq:eff_lagrangian}) is replaced by $H_d$. Both Higgs doublets participate in EWSB, their neutral components  develop vacuum expectation values, $\left< H_u^0 \right> = v_u$ and $\left< H_d^0 \right> = v_d$, with $\sqrt{v_u^2 + v_d^2} = v = 174$ GeV and  $\tan \beta \equiv v_u / v_d$. Thus, in the charged lepton mass matrix (\ref{eq:mass_matrix}), and Eqs. (\ref{eq:m_mu}) and (\ref{eq:m^LE}), $v$ is replaced by $v_d$.

The additional neutral and charged Higgs bosons, $H,\,A,\,H^\pm$, also contribute to $(g-2)_{\mu}$ and, assuming one scale of new physics $M_{L,E} \simeq m_{H,A,H^\pm}$,  the total contribution  is very well approximated by~\cite{Dermisek:2020cod}\cite{Dermisek:2021ajd}
\begin{equation}
\Delta a_{\mu}  \simeq  - \frac{1+\tan^2 \beta}{16\pi^{2}}  \frac{m_\mu m_\mu^{LE}}{v^2} , \quad m_\mu^{LE} \equiv \frac{\lambda_{L} \bar{\lambda} \lambda_{E}}{M_{L}M_{E}} v_d^3,
\label{eq:dela_2HDM}
\end{equation}
 where, to avoid confusion with the SM case, we included the proper definition of $m_\mu^{LE}$ in type-II 2HDM. For smaller Higgs masses compared to masses of new leptons, the results are almost unchanged. With increasing the Higgs masses, the $\tan^2 \beta$ enhanced contributions of heavy Higgses go to zero and the formula, with redefined $\lambda$-couplings that absorb $\cos \beta$, matches the SM result (\ref{eq:dela}).\footnote{To be more specific, when heavy Higgs masses are split from those of new leptons we obtain a similar formula with the replacement $\tan^{2}\beta\rightarrow H(x)\tan^{2}\beta$, where $H(x)=x (-7+8x-x^2-2(2+x)\ln(x))/(1-x)^{3}$ and $x=M_{L,E}^{2}/m_{H}^{2}$ (assuming $M_{L}=M_{E}$). It is very well approximated by $H(x)\simeq 1$ for any $x \geq 1$ and goes to zero as heavy Higgses are decoupled.}

The explanation of $\Delta a_{\mu}$ in the type-II 2HDM with $M_{L,E} \simeq m_{H,A,H^\pm}$ requires
\begin{equation}
m_\mu^{LE}/m_\mu = (-1.07 \pm 0.25 )/(1+\tan^2 \beta),
\label{eq:mLE-2HDM}
\end{equation}
and we see that, after fixing the effective lagrangian by muon mass and muon $g-2$, one parameter, $\tan \beta$, remains and it controls predictions for $h \to \mu^+\mu^-$,  $\mu^+\mu^- \to hh$, and  $\mu^+\mu^- \to hhh$.\footnote{Note that $\tan \beta$ interpolates, with proper redefinition of $\lambda$-couplings, between the SM ($\tan\beta = 0$) and models without mass mixing between the muon and vectorlike leptons ($\tan \beta = \infty$). Although both limits are not physical within 2HDM, this provides an insight to the $\tan \beta$ dependence of discussed results.}

\begin{figure}[t]
\includegraphics[scale=0.3]{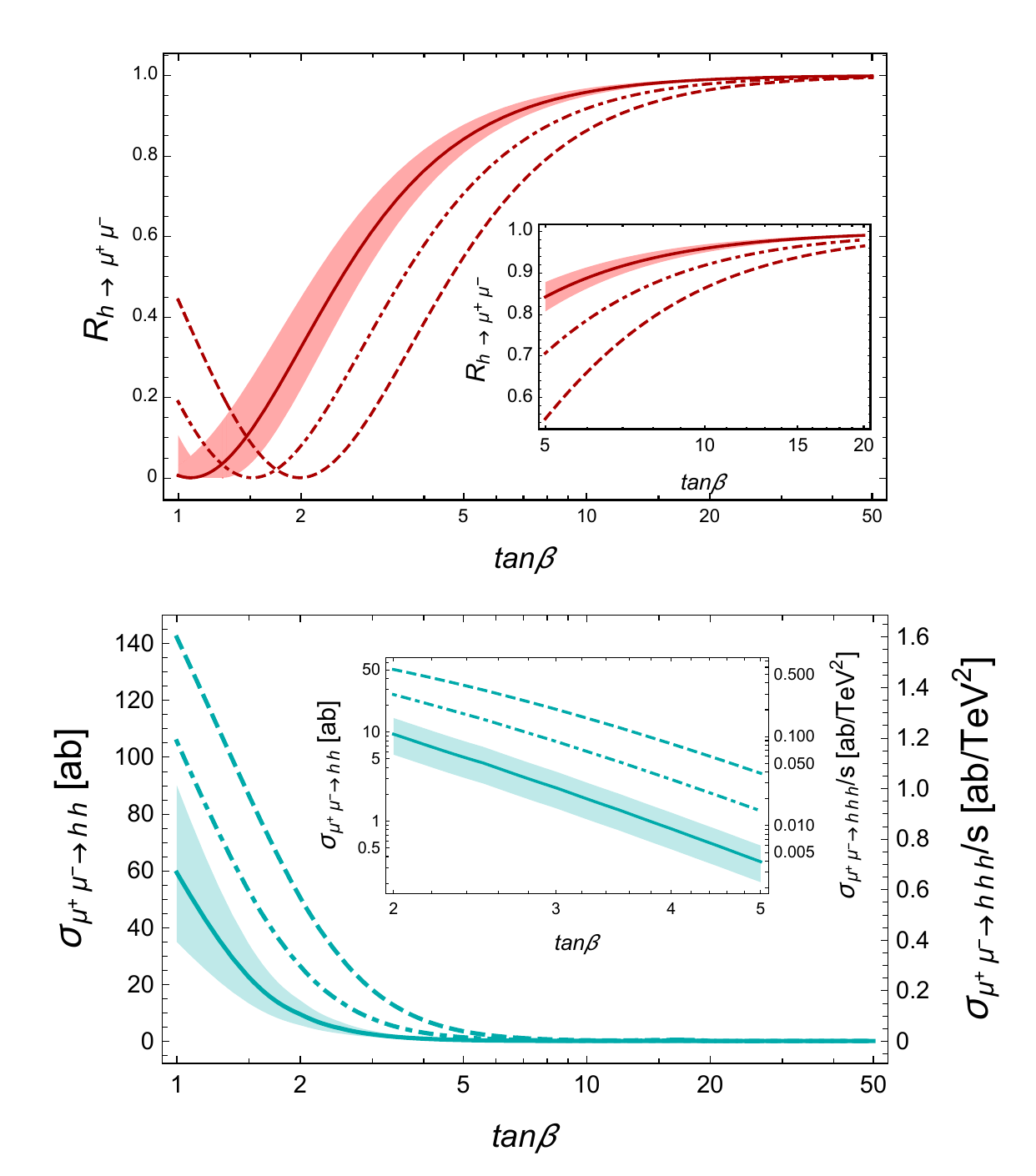}
\caption{$R_{h\to\mu^{+}\mu^{-}}$ (top), $\sigma_{\mu^{+}\mu^{-}\to h h }$ (bottom, left vertical axis) and $\sigma_{\mu^{+}\mu^{-}\to h h h}/s$  (bottom, right vertical axis) as a functions of $\tan\beta$ corresponding to the  central value of $\Delta a_\mu$ (solid lines) and its  one sigma range (shaded regions) from the effective Lagrangian in type-II 2HDM assuming $M_{L,E} \simeq m_{H,A,H^\pm}$. 
The dot-dashed and dashed lines are predictions corresponding to  the central value of $\Delta a_{\mu}$ for $m_{H,A,H^\pm} = 3\;\times M_{L,E}$ and $5\times M_{L,E}$. 
}
\label{fig:R_vs_tanb}
\end{figure}

The formulas for couplings describing interactions of the muon with the SM Higgs boson defined by Eq.~(\ref{eq:lagrangian_h}), namely the second equality in Eq.~(\ref{eq:lambda_h}) and Eqs.~(\ref{eq:lambda_hh})-(\ref{eq:lambda_hhh}) are the same with proper definitions of $m_\mu^{LE}$ for 2HDM given in Eq.~(\ref{eq:dela_2HDM}). The range of $R_{h\to \mu^+\mu^-}$ corresponding to one sigma range of $\Delta a_{\mu}$ straightforwardly follows and is plotted as a function of $\tan \beta$ in Fig.~\ref{fig:R_vs_tanb} (top). Note that $R_{h\to \mu^+\mu^-}$ drops to zero at small $\tan\beta $ as a result of cancellation in Eq.~(\ref{eq:R}). However, it should be  expected that for lighter new leptons, or when masses of Higgs bosons and new leptons are split,  Eq.~(\ref{eq:R}) is not a good approximation in this region, since small differences in model parameters could prevent perfect cancellation, as indicated by dot-dashed and dashed lines.  
Nevertheless, the measurement of  $h\to \mu^+\mu^-$ will constrain the range of $\tan\beta $, $M_{L,E}$ and $m_{H,A,H^\pm}$ consistent with $\Delta a_{\mu} $. The inset zooms in the range of  $\tan\beta $ where the deviation of $R_{h\to \mu^+\mu^-}$ from the SM expectation is comparable to the ultimate sensitivity of the LHC.

The common $\tan \beta$ dependence of $\mu^+\mu^- \to hh$ and $\mu^+\mu^- \to hhh$ total cross sections corresponding to  the central value and  one sigma range of $\Delta a_{\mu}$, assuming $M_{L,E} \simeq m_{H,A,H^\pm}$,  is plotted in Fig.~\ref{fig:R_vs_tanb} (bottom). Note that for $\mu^+\mu^- \to hhh$ we plot the total cross section divided by $s$ (right vertical axis). As $\tan\beta$ increases the SM Higgs boson contributes only a small fraction of $\Delta a_{\mu}$ and, correspondingly, the di-Higgs and tri-Higgs production rates drop as $\tan^4\beta$. For lighter Higgses compared to new leptons the predictions do not change significantly. However, with increasing $m_{H,A,H^\pm}$ compared to $M_{L,E}$  the contribution of the SM Higgs boson is increasing  that also reflects in increasing rates for $\mu^+\mu^- \to hh$ and $\mu^+\mu^- \to hhh$ (see dot-dashed and dashed lines).

\section{Discussion and Conclusions}

The di-Higgs and tri-Higgs productions present unique opportunities for a muon collider to test some of the simplest explanations of $\Delta a_\mu$ that do not require gauge or scalar mediators besides those already present in the SM. The $\mu^+\mu^- \to hh$ signal benefits from its production cross section being independent of $\sqrt{s}$ and thus a low energy machine is sufficient (note however, that the expected integrated luminosity scales with energy~\cite{Delahaye:2019omf}). For example, a $\sqrt{s} = 1$ TeV muon collider with 0.2 ab$^{-1}$ of integrated luminosity could see about 50 di-Higgs events if  the studied model with leptons with $SU(2)\times U(1)$ quantum numbers $\mathbf{2}_{-1/2}\oplus\mathbf{1}_{-1}$, copying the SM leptons, is the correct explanation of $\Delta a_\mu$.
At $\sqrt{s} = 3$ TeV with 1 ab$^{-1}$ of integrated luminosity a muon collider is expected to see about 30 tri-Higgs events in addition to about 240 di-Higgs events. Note, that  SM backgrounds for both these processes are negligible, see \cite{Chiesa:2020awd} for an interesting discussion of the challenging $hhh\bar \nu \nu$ production in the SM at a muon collider.

Other models with more exotic quantum numbers of new leptons will also lead to di-Higgs and tri-Higgs production in a similar way. The mass-enhanced contribution to $(g-2)_{\mu}$ is given by Eq.~(\ref{eq:dela}) multiplied by the corresponding $c$ factor which is~\cite{Kannike:2011ng}:  $c = 5$ for $\mathbf{2}_{-1/2}\oplus\mathbf{3}_{-1}$, $c=3$ for $\mathbf{2}_{-3/2}\oplus\mathbf{1}_{-1}$ or for $\mathbf{2}_{-3/2}\oplus\mathbf{3}_{-1}$, and $c=1$ for $\mathbf{2}_{-1/2}\oplus\;\mathbf{3}_{0}$. The resulting total cross sections are divided by $c^2$ compares to those presented in Fig.~\ref{fig:cross_sections_SM}. Thus 1 ab$^{-1}$ of integrated luminosity is sufficient to cover all possibilities using $\mu^+\mu^- \to hh$ and, more importantly, sufficient to distinguish between them (up to degeneracies).

In type-II 2HDM, the SM Higgs contributes significantly to  $\Delta a_\mu$ only at small $\tan\beta$  and, correspondingly, only in this region the cross sections for $\mu^+\mu^- \to hh$ and $\mu^+\mu^- \to hhh$ are large. However, this is exactly the region that allows for the heaviest, even 10s of TeV, new leptons and Higgs bosons to explain $\Delta a_\mu$~\cite{Dermisek:2020cod, Dermisek:2021ajd}.  Not seeing the expected rates  would  provide constraints on $\tan \beta$ and hierarchies between the leptons and new Higgses, but more importantly, it would  constrain the allowed parameter space to significantly lower masses, effects of which might be accessible elsewhere. Furthermore, with increasing $\tan\beta$ the contributions of heavy Higgses dominate $\Delta a_\mu$ and, correspondingly, large rates for $\mu^+\mu^- \to HH,\, AA,\, H^+H^-$ and $\mu^+\mu^- \to hHH,\, hAA,\, hH^+H^-$ are predicted, similar to $hh$ and $hhh$ in Fig.~\ref{fig:cross_sections_SM}, for $\sqrt{s}$ above the threshold. 

In models with mass enhancement in the contribution of new leptons but mediated by a new scalar $S$ not participating in EWSB, for summary and references see, for example, Ref.~\cite{Capdevilla:2021rwo}, similar rates are expected  for $SS$ and $hSS$ production  above the threshold. Analogous reasoning also applies to new gauge boson mediators. Indeed, already in the main model discussed here, there are similar signatures corresponding to $Z$ and $W$ loops contributing  to $\Delta a_\mu$, namely $\mu^+\mu^- \to ZZ,\, W^+\,W^-$ and $\mu^+\mu^- \to hZZ,\, hW^+\,W^-$ with comparable rates to $hh$ and $hhh$ (but also with much larger SM backgrounds). For a general discussion of new physics at muon colliders see~\cite{AlAli:2021let}. SM backgrounds for some of the mentioned processes here and related useful discussion can be found in~\cite{Han:2021lnp}.

If the $(g-2)_{\mu}$ anomaly persists, not observing the expected rates for $\mu^+\mu^- \to hh$ and $\mu^+\mu^- \to hhh$ at a muon collider would be a clear evidence for the existence of new forces, gauge bosons or scalars, beyond those in the SM. The confirmation of such scenarios, might require a $\sim30$ TeV muon collider, while still relying on additional assumptions  like naturalness~\cite{Capdevilla:2020qel}\cite{Capdevilla:2021rwo} or  the ability to detect the related $\mu^+ \mu^- \to h\gamma$ signal on a huge SM background~\cite{Buttazzo:2020eyl, Yin:2020afe}, or a ${\cal O}(100)$ TeV muon collider to directly produce new leptons.



\acknowledgments
The work of R.D. was supported in part by the U.S. Department of Energy under Award No. {DE}-SC0010120. TRIUMF receives federal funding via a contribution agreement with the National Research Council of Canada.

\appendix

\section{$\sigma_{\mu^{+} \mu^{-} \rightarrow h h}$ and $\sigma_{\mu^{+} \mu^{-} \rightarrow h h h}$ cross sections}

We first present the full calculation for $\sigma_{\mu^{+} \mu^{-} \rightarrow h h}$ as a function of $\sqrt{s}$ and $m_{e_{4,5}}$. Defining the unitless variable $x_a = \sqrt{s^2 - 4 m_h^2 s}/(2 m_{e_a}^2 + s - 2m_h^2)$, where $m_h$ is the SM Higgs boson mass, and neglecting the muon mass we obtain 
\begin{widetext}
\begin{equation}
\begin{aligned}
   \sigma_{\mu^{+} \mu^{-} \rightarrow h h} = \frac{1}{64 \pi}
   \left(\sqrt{1 - \frac{4 m_h^2}{s}} \right) & \sum_{a,b = 4, 5} \left[ m_{e_a} m_{e_b} \textrm{Re} \left[\lambda_{\mu e_a}^h \lambda_{e_a \mu}^h \lambda_{\mu e_b}^{h *} \lambda_{e_b \mu}^{h *} \right] A(s; x_a, x_b) \right. \\
   & \left. + \frac{1}{8} \left(|\lambda_{\mu e_a}^h|^2 |\lambda_{\mu e_b}^h|^2 + |\lambda_{e_a \mu}^h|^2 |\lambda_{e_b \mu}^h|^2 \right) (s - 4m_h^2)B(s; x_a, x_b) \right],
\label{eq:fullsigma_hh}
\end{aligned}
\end{equation}
where
\begin{eqnarray}
\begin{aligned}
A(s; x_a, x_b) = \frac{4 x_a x_b}{s^2 - 4m_h^2 s} \left(\frac{1}{x_a^2 - x_b^2} \right) \left[x_a \textrm{arctanh}(x_a) - x_b \textrm{arctanh}(x_b) \right], \\
\end{aligned}
\end{eqnarray}
and
\begin{eqnarray}
\begin{aligned}
B(s; x_a, x_b) = \frac{4 x_a x_b}{s^2 - 4m_h^2 s} \left[\frac{x_b}{x_a^2 - x_b^2} \left(1 - \frac{1}{x_a^2} \right) \textrm{arctanh}(x_a) - \frac{x_a}{x_a^2 - x_b^2} \left(1 - \frac{1}{x_b^2} \right) \textrm{arctanh}(x_b) - \frac{1}{x_a x_b} \right].
\end{aligned}
\end{eqnarray}
The couplings $\lambda_{\mu e_{a,b}}^h$ and $\lambda_{e_{a,b}\mu}^h$ of leptons to the SM Higgs boson in mass eigenstate basis can be found in Appendix A2 of~\cite{Dermisek:2021ajd}.
Note that in the limit $x_{a,b} \ll 1$
\begin{eqnarray}
A(s; x_a, x_b) &\simeq& \frac{1}{m_{e_a}^2 m_{e_b}^2} \left[ 1 + \frac{1}{3} \left( x_a^2 + x_b^2 \right) \right],\\
B(s; x_a, x_b) &\simeq& \frac{2}{15} \frac{x_a x_b}{m_{e_a}^2 m_{e_b}^2}.
\end{eqnarray}
Keeping only the $\mathcal{O}(1)$ term, the $\mu\mu \rightarrow h h$ cross-section becomes
\begin{eqnarray}
\begin{aligned}
\sigma_{\mu^{+} \mu^{-} \rightarrow h h} \simeq \frac{1}{64 \pi} \left(\sqrt{1 - \frac{4 m_h^2}{s}} \right) \sum_{a,b = 4, 5} \frac{1}{m_{e_a} m_{e_b}} \textrm{Re} \left[\lambda_{\mu e_a}^h \lambda_{e_a \mu}^h \lambda_{\mu e_b}^{h *} \lambda_{e_b \mu}^{h *} \right]. \\
\end{aligned}
\end{eqnarray}
Summing over new fermions, we find 
\begin{eqnarray}
\begin{aligned}
\sigma_{\mu^{+} \mu^{-} \rightarrow h h} \simeq \frac{1}{64 \pi} \left(\sqrt{1 - \frac{4 m_h^2}{s}} \right) \Big| \frac{ \lambda_{\mu L}^h \lambda_{L \mu}^h}{M_L} + \frac{ \lambda_{\mu E}^h \lambda_{E \mu}^h}{M_E}  \Big|^2 = \frac{9}{64 \pi} \left(\sqrt{1 - \frac{4 m_h^2}{s}} \right) \left( \frac{m_{\mu}^{LE}}{v^2}\right)^2 \label{eq:sigma_hh_EFT},
\end{aligned}
\end{eqnarray}
where we have used the approximate formulas for couplings provided in Appendix A4 of~\cite{Dermisek:2021ajd}. This matches the effective field theory result (shown in Eq.~\ref{eq:EFT_xsections_1} in the limit $m_{h}\rightarrow 0$).

The exact result for $\mu \mu \rightarrow h h h$ cross-section is quite lengthy and not particularly revealing. However, in the limit of heavy lepton masses (and neglecting the Higgs mass) the result highly simplifies to
\begin{eqnarray}
\begin{aligned}
\sigma_{\mu^+ \mu^- \rightarrow h h h} \simeq \frac{3 s}{8192 \pi^3} \sum_{a,b,c,d = 4, 5} \frac{ \textrm{Re} \left[ \lambda_{e_a \mu}^h \lambda_{\mu e_b}^{h} \lambda_{e_b e_a}^h \lambda_{e_c \mu}^{h *} \lambda_{\mu e_d}^{h *} \lambda_{e_d e_c}^{h *} + \lambda_{\mu e_a}^{h *} \lambda_{e_b \mu}^{h *} \lambda_{e_a e_b}^{h *} \lambda_{\mu e_c}^{h} \lambda_{e_d \mu}^{h} \lambda_{e_c e_d}^{h}  \right]}{m_{e_a} m_{e_b} m_{e_c} m_{e_d}}.
\end{aligned}
\end{eqnarray}
Summing over new fermions, we find
\begin{eqnarray}
\begin{aligned}
\sigma_{\mu^+ \mu^- \rightarrow h h h} \simeq \frac{3 s}{4096 \pi^3} \Big| \frac{ \lambda_{L \mu}^h \lambda_{L L}^h \lambda_{\mu L}^h}{M_L^2} + \frac{\lambda_{L \mu}^h \lambda_{E L}^h \lambda_{\mu E}^h}{M_L M_E} + \frac{ \lambda_{\mu L}^h \lambda_{L E}^h \lambda_{E \mu}^h}{M_L M_E} + \frac{\lambda_{E \mu}^h \lambda_{E E}^h \lambda_{\mu E}^h}{M_E^2} \Big|^2 \simeq \frac{3}{4096 \pi^3} \left( \frac{m_{\mu}^{LE}}{v^3} \right)^2 s,
\end{aligned}
\end{eqnarray}
where in the last equality we have used approximate formulas for couplings found in Appendix A4 of~\cite{Dermisek:2021ajd}. This matches the effective field theory result found in Eq.~\ref{eq:EFT_xsections_2}.\\
\end{widetext}
%





\end{document}